\documentclass[dvips]{acta}
\usepackage{supertabular,lscape,epsfig}
\usepackage{amssymb}
\usepackage{amsmath}
\usepackage{float}
\usepackage{statmath}
\mathchardef\mhyphen="2D
\usepackage{placeins}
\DeclareSymbolFont{ppa}{OT1}{ppl}{m}{it}
\DeclareMathSymbol{\vv}{\mathalpha}{ppa}{'166}

\SetPages{0}{0}

\SetVol{75}{2025}
\newcommand{\doi}{~}

\begin{document}
\newcommand\pvalue{\mathop{p\mhyphen {\rm value}}}
\newcommand{\TabApp}[2]{\begin{center}\parbox[t]{#1}{\centerline{
  {\bf Appendix}}
  \vskip2mm
  \centerline{\small {\spaceskip 2pt plus 1pt minus 1pt T a b l e}
\refstepcounter{table}\thetable}
  \vskip2mm
  \centerline{\footnotesize #2}}
  \vskip3mm
\end{center}}

\newcommand{\TabCapp}[2]{\begin{center}\parbox[t]{#1}{\centerline{
  \small {\spaceskip 2pt plus 1pt minus 1pt T a b l e}
  \refstepcounter{table}\thetable}
  \vskip2mm
  \centerline{\footnotesize #2}}
  \vskip3mm
\end{center}}

\newcommand{\TTabCap}[3]{\begin{center}\parbox[t]{#1}{\centerline{
  \small {\spaceskip 2pt plus 1pt minus 1pt T a b l e}
  \refstepcounter{table}\thetable}
  \vskip2mm
  \centerline{\footnotesize #2}
  \centerline{\footnotesize #3}}
  \vskip1mm
\end{center}}

\newcommand{\MakeTableApp}[4]{\begin{table}[p]\TabApp{#2}{#3}
  \begin{center} \TableFont \begin{tabular}{#1} #4 
  \end{tabular}\end{center}\end{table}}

\newcommand{\MakeTableSepp}[4]{\begin{table}[p]\TabCapp{#2}{#3}
  \begin{center} \TableFont \begin{tabular}{#1} #4 
  \end{tabular}\end{center}\end{table}}

\newcommand{\MakeTableee}[4]{\begin{table}[htb]\TabCapp{#2}{#3}
  \begin{center} \TableFont \begin{tabular}{#1} #4
  \end{tabular}\end{center}\end{table}}

\newcommand{\MakeTablee}[5]{\begin{table}[htb]\TTabCap{#2}{#3}{#4}
  \begin{center} \TableFont \begin{tabular}{#1} #5 
  \end{tabular}\end{center}\end{table}}

\newcommand{\MakeTableH}[4]{\begin{table}[H]\TabCap{#2}{#3}
  \begin{center} \TableFont \begin{tabular}{#1} #4 
  \end{tabular}\end{center}\end{table}}

\newcommand{\MakeTableHH}[4]{\begin{table}[H]\TabCapp{#2}{#3}
  \begin{center} \TableFont \begin{tabular}{#1} #4 
  \end{tabular}\end{center}\end{table}}
\newfont{\bb}{ptmbi8t at 12pt}
\newfont{\bbb}{cmbxti10}
\newfont{\bbbb}{cmbxti10 at 9pt}
\newcommand{\uprule}{\rule{0pt}{2.5ex}}
\newcommand{\douprule}{\rule[-2ex]{0pt}{4.5ex}}
\newcommand{\dorule}{\rule[-2ex]{0pt}{2ex}}
\def\thefootnote{\fnsymbol{footnote}}
\begin{Titlepage}

\Title{The Ultimate {\it I}-band Calibration of the TRGB Standard Candle}
\Author{
A.~~U~d~a~l~s~k~i$^1$,~~ D.\,M.~~S~k~o~w~r~o~n$^1$,~~
J.~~S~k~o~w~r~o~n$^1$, ~~
M.\,K.~~S~z~y~m~a~ń~s~k~i$^1$,~~
I.~~S~o~s~z~y~ń~s~k~i$^1$,~~
P.~~P~i~e~t~r~u~k~o~w~i~c~z$^1$,~~
P.~~M~r~ó~z$^1$,\\
R.~~P~o~l~e~s~k~i$^1$,~~
S.~~K~o~z~ł~o~w~s~k~i$^1$,~~
K.~~U~l~a~c~z~y~k$^{2,1}$,~~
K.~~R~y~b~i~c~k~i$^{3,1}$,\\
P.~~I~w~a~n~e~k$^1$,~~
M.~~G~r~o~m~a~d~z~k~i$^1$,~~
M.~~W~r~o~n~a$^{4,1}$,\\
M.~~R~a~t~a~j~c~z~a~k$^1$,~~
and~~~M.\,J.~~M~r~ó~z$^1$
}
{$^1$Astronomical Observatory, University of Warsaw, Al.~Ujazdowskie~4, 00-478~Warszawa, Poland\\
$^2$Department of Physics, University of Warwick, Gibbet Hill Road, Coventry, CV4~7AL,~UK\\
$^3$Department of Particle Physics and Astrophysics, Weizmann Institute of Science, Rehovot 76100, Israel\\
$^4$Department of Astrophysics and Planetary Science, Villanova University,\\ 800 East Lancaster Avenue, Villanova, PA 19085, USA}
\end{Titlepage}

\Abstract{We present the ultimate {\it I}-band calibration of the tip
of the red giant branch (TRGB) standard candle. Our calibration is
based on photometry from the outer parts of the Large Magellanic
Cloud, $2\zdot\arcd75<r<6\zdot\arcd5$ from the center, collected
during the OGLE-IV phase of the Optical Gravitational Lensing
Experiment. Outer regions of the LMC have large advantages compared to
the previous attempts of the TRGB calibrations using the red giants
from the central parts of this galaxy. The interstellar reddening in
these regions is much lower and more uniform, stellar crowding is
lower and the outer parts of the LMC can be accurately described as a
flat disk within the reasonable distance from the LMC center. The
number of red giants in the upper part of the red giant branch in our
LMC region is large, $\approx 140\,000$, making it possible to
determine of the tip magnitude with high accuracy.

Our ultimate {\it I}-band calibration of the TRGB is:
$M_{I,TRGB}=-4.022 \pm 0.006~{\rm(stat.)} \pm 0.033~{\rm(syst.)}$
mag. We also provide its values for different techniques of the
determination of the tip magnitude. The accuracy of our calibration is
mostly limited by the accuracy of the distance to the LMC ($\approx
1\%$) and can be improved in the future.

We test our calibration by comparing it with the TRGB in the Small
Magellanic Cloud and NGC~4258, \ie the galaxies with precise geometric
distance determination, and find excellent agreement. Finally, we
refine the main determinations of the Hubble constant, $H_0$, with the
TRGB using our new calibration of the {\it I}-band TRGB
brightness.}{~}

\Section{Introduction}

Since many decades, red giant (RG) stars have been used in attempts of
the distance determination to objects hosting them. However, the first
modern proposition of using the {\it I}-band brightness of the tip of
the red giant branch (TRGB) on the color magnitude diagram (CMD) as a
potential standard candle came and was described in the early 1990s by
Lee, Freedman and Madore (1993). It was suggested that the sharp
boundary in the density of the red giants located at the top of the
red giant branch (RGB) remains constant in the {\it I}-band with high
accuracy. Theoretical studies of evolutionary models of giants
confirmed this feature (\cf Freedman \etal 2020) making the TRGB a
promising distance indicator. Since the 1990s, the TRGB technique has
been applied to the determination of distances to many Local Group
galaxies and located beyond it -- both from the ground and from space
with the HST (\cf Freedman \etal 2019, Kim \etal 2020), and recently,
the JWST (\cf Freedman \etal 2025). It should be noted that the TRGB
is relatively bright, so this technique makes it possible to reach
galaxies that are beyond the range of other stellar standard candles
like RR Lyrae stars or even Cepheids. It is also easier to apply and
requires less precious -- often satellite -- observing time than using
variable stars. Thus, the TRGB technique has become a vital
alternative to, for example, Cepheids.

The development of the TRGB method proceeded in two directions. First,
techniques for the homogeneous and precise determination of TRGB
magnitudes were developed (\eg Lee \etal 1993, Makarov \etal
2006). This is a non-trivial task, because while it is relatively easy
to visually spot the presence of a density drop at the top of the RGB,
the accurate numerical determination of the magnitude of this boundary
requires a sufficiently well populated red giant branch and advanced
numerical methods.  Second, calibrations of the TRGB were attempted
based mostly on red giants from the Large Magellanic Cloud
(LMC). However, the results of these calibrations were far from being
satisfactory, showing significant scatter in the absolute {\it I}-band
magnitude of the TRGB (\eg Table~1, Freedman 2021). Recently, the
NGC~4258 galaxy -- with a 1\% accurate distance determination {\it
via} water masers -- was also used for calibrating the absolute
magnitude of the TRGB (Jang \etal 2021, Scolnic \etal 2023), also in
the F090W filter of the JWST space telescope (Anand \etal 2024). It is
expected that the JWST will provide many new TRGB measurements in
distant galaxies in the coming years. More details and comprehensive
information on the TRGB distance determination technique can be found
in the review article by Li and Beaton (2024).

The TRGB distance determinations should play a crucial role in resolving
or maintaining the discrepancy between the Hubble constant value
measured locally using standard candles, \eg
$H_0=73.0\pm1.0$~km/s/Mpc from Cepheids (Riess \etal 2022) and that
obtained by the Planck mission from the young Universe
($H_0=67.4\pm0.5$~km/s/Mpc, Planck collaboration \etal 2020). This
issue is known now as the Hubble tension and is considered one of the
most pressing challenges in modern astrophysics.

TRGB distances to local galaxies may allow for the determination
of the local Hubble constant independently from Cepheids. Yet, 
recent results have been ambiguous. Freedman (2021) obtained
$H_0=69.8 \pm0.6 {\rm (stat.)} \pm1.6 {\rm (syst.)}$ km/s/Mpc -- significantly
different from the Cepheid value and almost in the middle of the Hubble tension
gap. However, the uncertainties were still too large to draw definitive
conclusions. On the other hand, Anand \etal (2022) and Scolnic \etal
(2023) obtained different results from TRGB distances:
$H_0=71.5\pm1.5$~km/s/Mpc and $H_0=73.22\pm2.06$~km/s/Mpc,
respectively -- in a much better agreement with the local Cepheid value,
thereby sustaining the significance of the Hubble tension.  This
local discrepancy in $H_0$ values between Cepheids and TRGBs is highly
uncomfortable and is referred to as as the local Hubble tension.

One of the most crucial steps of the TRGB technique is the reliable
calibration of the {\it I}-band absolute magnitude of the TRGB. All
the most advanced recent calibration attempts (Jang and Lee 2017,
Yuan \etal 2019, Freedman \etal 2020, Hoyt 2023, Anderson \etal 2024)
used photometric data from the LMC collected
during the third phase of the OGLE project -- OGLE-III (Udalski
2003a).  Unfortunately, OGLE-III observed only the central parts of the LMC.
This makes the input data for calibration far from
being optimal, as it is well known that the central regions of the
LMC have high stellar density and relatively large, non-homogeneous
interstellar reddening. Moreover, the central regions of
the LMC are known to host several structures, which additionally make
the calibrations harder and more prone to systematic errors.

Here, we undertake a new approach to obtain the ultimate, most precise
calibration of the {\it I}-band absolute magnitude of the TRGB.  We
base our calibration on a completely new photometric data set --
extensive photometry of the LMC collected during the fourth phase of
the OGLE survey -- OGLE-IV (Udalski, Szyma{\'n}ski and Szyma{\'n}ski
2015). Because the OGLE-IV observations cover entire LMC we decided to
use only the outer regions of this galaxy in our TRGB calibration 
attempts. In these regions the interstellar reddening is low, stellar
density is moderate and the structure of the galaxy can be accurately
modeled as a flat disk. Thus, we eliminate all the drawbacks of
previous calibration attempts.

Additionally, in our calibration we use the most precise maps of
interstellar reddening in the Magellanic Clouds derived from red clump
stars by Skowron \etal (2021). These maps were calculated based on the
same OGLE-IV photometric data set so our photometric analysis is fully
homogeneous.

Our calibration is carefully tested on galaxies for which geometric
distance determinations exist with a percent-level accuracy: the Small
Magellanic Cloud (SMC) and NGC~4258. Finally, we use our calibration
of the {\it I}-band absolute magnitude of the TRGB to refine recent
TRGB-based determinations of the Hubble constant.

\Section{Deep Photometric Maps of OGLE-IV}

Deep OGLE-IV photometric maps
contain mean {\it V-} and {\it I-}band photometry of
objects detected on the deep OGLE-IV reference images. The latter were
constructed by stacking up to one hundred best-seeing individual
images in the {\it I}-band and a dozen or so in the {\it V}-band -- in
the same manner as the regular OGLE-IV reference images (typically an
average of 5--8 individual frames) used for the Difference Image
Analysis (DIA, Wo{\'z}niak 2000) in the standard OGLE photometric
pipeline (Udalski 2003a). Effectively, the deep OGLE-IV images reach the
depth of individual exposures taken with 3--4 m class telescopes with 
typical OGLE exposure times (100--150 sec.) The effective seeing
of these deep images is of the order of
1\zdot\arcs1--1\zdot\arcs2. The first image in the stack is the same
in both the deep and standard OGLE reference images, which ensures
the exact same astrometric grid and the DIA photometry zero
level. Thus, both the deep and standard pipeline OGLE photometries
are on the same instrumental photometric system.

The PSF photometry with the DoPhot photometry program (Schechter,
Mateo and Sahu 1993) was applied to a deep image of each subfield of
the OGLE field and then converted to the DIA photometry instrumental
system. Each subfield corresponds to one the 32 CCDs of the OGLE-IV
mosaic camera. Each deep reference image is, in fact, a pixel average of
many individual images transformed to the same astrometric grid. Thus,
the deep photometry represents the mean brightness of objects --
averaged over a significant span of time, typically about five years.

In the next step, the stellar lists with the deep instrumental
photometry in the DIA system for the {\it V}- and {\it I}-band of a
given subfield were cross-matched using {\it V} $\Leftrightarrow$
{\it I} astrometric grid transformations, forming the instrumental deep
OGLE maps. Finally, the instrumental deep photometry was tied to the
standard $VI$ Johnson-Cousin system using the same calibration
procedures as for regular OGLE photometry (Udalski, Szyma{\'n}ski and
Szyma{\'n}ski 2015), since both are on the same instrumental system.

As expected, both deep and standard OGLE photometric maps are very
consistent.  The average difference in photometry for thousands of
stars from the deep and standard OGLE maps in the {\it V}- and {\it
I}-bands is at the 1--2 mmag level. On the other hand, the accuracy of
the absolute $VI$ calibration is at the 10--15 mmag level and 
depends on the stellar density of the field. In the fields analyzed
here, which are located farther than 2\zdot\arcd75 from the LMC center
and are of moderate stellar density, it is even below the 10 mmag level.

Fig.~1 presents the color-magnitude diagram (CMD) of the one of the
LMC OGLE fields analyzed in this study, LMC513, to demonstrate the
range and quality of the OGLE-IV deep photometric maps.

\begin{figure}[htb]
\centerline{\includegraphics[width=12cm]{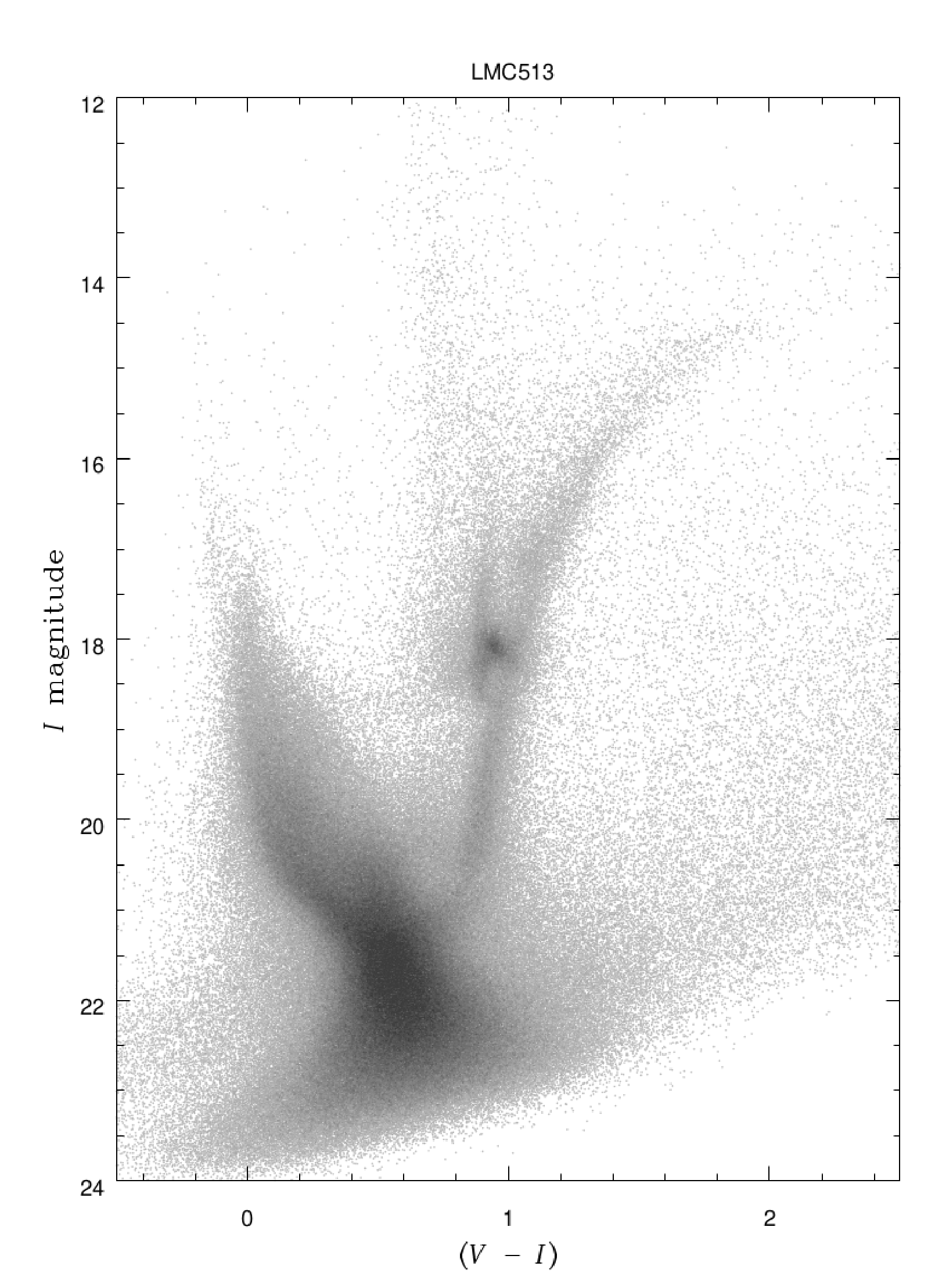}}
\vskip 12pt
\FigCap{Deep OGLE CMD of LMC513 -- one of 269 OGLE-IV fields in the
LMC. Darker colors indicate areas of higher density of the points.}
\end{figure}

\Section{Reddening Maps}

The interstellar reddening in the LMC is relatively low. Nevertheless,
there are regions, especially in the central parts of this galaxy, where it
is significantly higher. For example, in the star forming regions around the
Tarantula nebula it is very high and exhibits quite non-standard
properties. The determination of the reddening across the entire galaxy
has been a subject of many studies (\eg Subramaniam 2005, Pejcha and Stanek
2009, Haschke \etal 2011, Tatton \etal 2013, Inno \etal 2016,
Choi \etal 2018, Joshi and Panchal 2019, G{\'o}rski \etal 2020 and
others).

At the same time, information on interstellar reddening is crucial for the precise
determination of the {\it I}-band magnitude of the TRGB, and the uncertainty
resulting from this parameter is one of the main obstacles in obtaining
a commonly accepted accuracy of the TRGB calibration.
Therefore, in our study we use the most extensive and detailed maps of
interstellar reddening in the Magellanic Clouds, derived from 
OGLE-IV photometry by Skowron \etal (2021).

The Skowron \etal (2021) maps are based on red clump stars. The
intrinsic $(V-I)_0$ color of the red clump serves as the reference
point for the $E(V-I)$ reddening measurements along different
lines-of-sights.  The population of the red clump stars in the
Magellanic Clouds is large enough that the $(V-I)$ color, and thus
$E(V-I)$, can be derived with high accuracy in the large regions of
both Magellanic Clouds. The calibration of the $(V-I)_0$ reference
point was carried out very carefully. Since the intrinsic color of the
red clump, $(V-I)_0$, depends slightly on the metallicity of the
environment, the calibration accounts for the small metallicity
gradient observed in both Clouds. The maps of Skowron \etal (2021)
have a very good spatial resolution: in the central parts it is
$1\zdot\arcm7 \times 1\zdot\arcm7$ decreasing to $27\arcm \times
27\arcm$ in the outskirts, where the maps smoothly join with the all
sky reddening map of Schelgel, Finkbeiner and Davis (SFD, 1998) after
the correction by Schlafly and Finkbeiner (2011).

\Section{TRGB Calibration}

\subsection{The LMC Sample}

Although the OGLE project contributed to the TRGB studies as early as
at the OGLE-II phase (Udalski 2000), the main previous attempts of
TRGB calibration based on OGLE photometry relied on data from the
central regions of the LMC collected during the OGLE-III phase (Jang
and Lee 2017, Yuan \etal 2019, Freedman \etal 2020, Hoyt 2023,
Anderson \etal 2024). As already mentioned in Section~1, these regions
are not ideally suited for TRGB calibration, due to high stellar
density and relatively large and non-uniform interstellar
reddening. Therefore, we decided to use the outer, peripheral regions
of this galaxy to perform the most precise calibration based on the
OGLE photometry collected during the fourth phase of the OGLE project.
The extinction in this part of the LMC is lower and more homogeneous
than in the central parts (Skowron \etal 2021).

The LMC, like every galaxy, has a 3D structure and contains different
stellar populations. In the central part, the characteristic feature
is a stellar bar containing a young population, including numerous
classical Cepheids which precisely define its shape
(Jacyszyn-Dobrzeniecka \etal 2016). The outer regions form a rather
flat stellar disk, also containing some young population -- classical
Cepheids -- but predominately made up of older, intermediate age
stars, including RG and TRGB objects. The RG stars in the LMC disk can
be well modeled by a flat disk using red clump giants (Choi \etal
2018, Saroon and Subramanian 2022), parametrized by the inclination
angle, $i$, between the disk and sky plane, and the position angle,
$PA$, between the North direction in the sky and the line of nodes,
measured from North toward East. Recent estimates of these parameters
for the red clump stars disk are: ($i$, $PA$) = (27\zdot\arcd81,
146\zdot\arcd37, Choi \etal 2018) and ($i$, $PA$) = (23\zdot\arcd26,
160\zdot\arcd43, Saroon and Subramanian 2022).

There exists some evidences that the stellar disk of the LMC is warped
toward the SMC (Choi \etal 2018, Saroon and Subramanian 2022) likely
due to interactions with this galaxy. However, the effects of disk
bending due to the warp become noticeable at about 5--6 kpc from the
LMC center. On the other hand, these distant regions of the LMC are
relatively poorly populated by stars so the effects of warping on the
general photometric properties of the disk are rather small.

Additional important parameters of the LMC disk are the equatorial
coordinates (RA, DEC) of its center on the sky. The position of the
LMC center has been debated in recent decades. The photometric center
of the LMC is often assumed to be at (RA, DEC) = (82\zdot\arcd25,
$-69\zdot\arcd5$; van der Marel and Cioni 2001). Further
determinations based on HST astrometry placed the LMC center at (RA,
DEC) = (79\zdot\arcd88, $-69\zdot\arcd6$) for the older population
(van der Marel and Kallivayalil 2014). The most recent determination
by the Gaia team using DR3 astrometry and kinematics locates it at
(RA, DEC) = (81\zdot\arcd07, $-69\zdot\arcd41$; Gaia
Collaboration \etal 2021), although it is at a slightly different
position when the central stars are not included. Since we will
analyze only the outer LMC disk, limited to $r>2\zdot\arcd75$ from the
center, we adopted as our default center position the average
coordinates from the two bottom lines of Table~5 in Gaia
Collaboration \etal (2021), \ie (RA, DEC) = (81\zdot\arcd465,
$-69\zdot\arcd515$).

For the TRGB calibration, we limited our analysis to the regions of
the LMC located outside the inner galaxy, defined as closer than
2\zdot\arcd75 from the center. This approach minimizes the impact of
the non-homogeneous features in the inner LMC regions (central bar)
and the intrinsic extinction of the LMC, which is significantly higher
in the central parts. The upper limit of $6\zdot\arcd5$, corresponding
to the distance of 5.7 kpc, was chosen to minimize potential bias from
the LMC disk warp (Choi \etal 2018, Saroon and Subramanian 2022).

\subsection{Initial Determination of the TRGB Brightness, $I_0^{TRGB}$}

We divided the disk of the LMC into 12 sectors, as described below.
We first defined two rings around the center -- A:
$2\zdot\arcd75<r<4\zdot\arcd25$ from the center and B:
$4\zdot\arcd25<r<6\zdot\arcd5$.  Each ring was further divided into six sectors
(1--6), each 60\arcd wide.  The first sector starts at the position
angle $PA_0=130\arcd$, which roughly corresponds to the LMC disk's
line of nodes.  To check potential systematics that could arise from
this choice, we repeated the further analysis using also a starting
position angle of $PA_0=150\arcd$. Thus, six sectors (A1--A3 and
B1--B3) cover the part of the disk which is closer to us, while the
remaining six (A4--A6 and B4--B6) cover the more distant side. In
total, we individually analyze RGB stars in twelve sectors.

\setcounter{figure}{1}

\begin{figure}[htb]
\centerline{\includegraphics[width=13cm, bb=50 250 570 680]{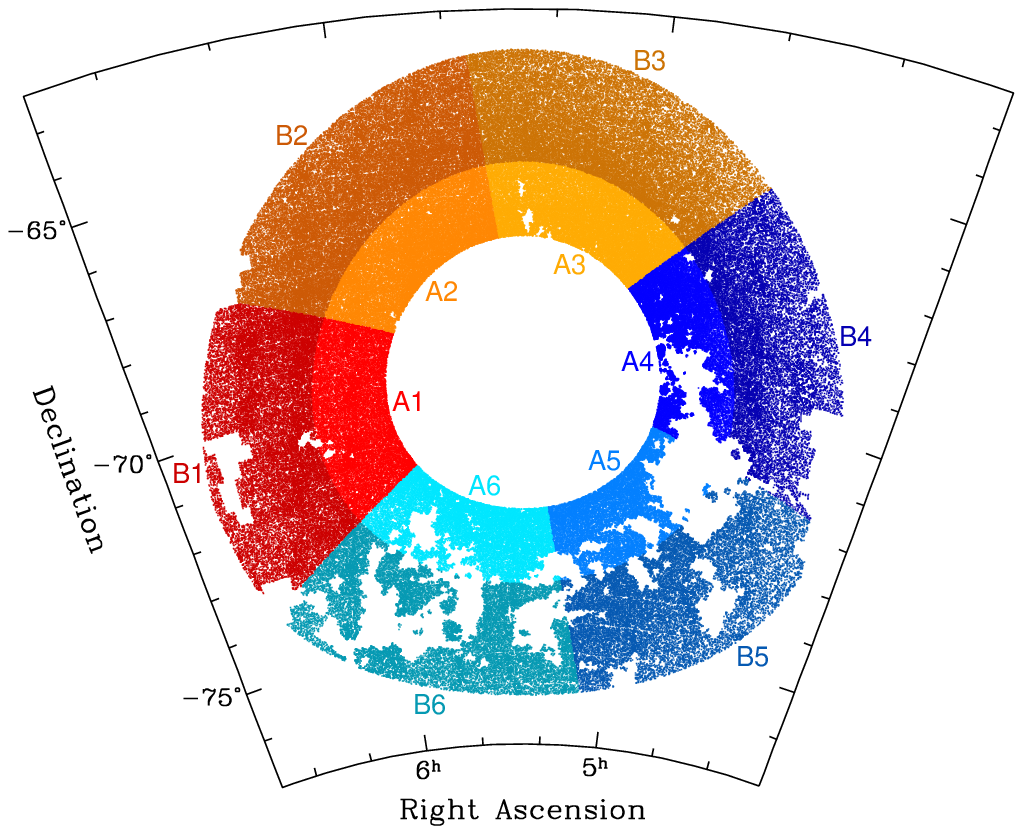}}
\FigCap{The disk of the LMC analyzed for the TRGB calibration. Different
colors show the segmentation scheme used for the initial determination of
$I_0^{TRGB}$. White spots are the regions, where the reddening is
higher, \ie $E(V-I)>0.15$~mag.}
\end{figure}

To minimize errors related to the reddening uncertainty, we excluded areas
with higher reddening: $E(V-I) \ge 0.15$~mag within each sector. The 
sectors contain from $\approx 8500$ to $\approx17\,500$ upper RGB stars
brighter than $I=16.0$~mag and redder than $V-I=0.9$~mag, which is a
sufficient number for the initial determination of the TRGB magnitude.
Fig.~2 shows a schematic view of our segmentation of the LMC disk on the sky.

We treated each RGB star in a sector individually. First, we corrected
its magnitude for the average interstellar extinction along the star's
line-of-sight using the reddening maps of Skowron \etal (2021) to
bring it to its intrinsic brightness. We followed the map prescription
and used the red clump $E(V-I)$ in the regions of the LMC where the
spatial resolution of the maps was high (better than 7\arcm), and the
SFD $E(V-I)$, also provided with the maps, in the remaining
regions. Then the total {\it I}-band extinction toward the star was
calculated as $A_I=R_I*E(V-I)$.

Conversion from reddening to the {\it I}-band extinction is somewhat
uncertain, as it is known that the coefficient $R_V$, describing the
extinction curve, varies in different regions of the Magellanic Clouds
(\cf Fig. 5 in Zhang and Green 2025). In the highly reddened OGLE-II
fields of the LMC, located roughly on the node line, about one degree
from its center (RA, DEC) = (84\zdot\arcd4, $-70\zdot\arcd22$), the
observed red clump shift due to reddening corresponds to $R_I=1.44$
(Udalski 2003b). However, according to the map of Zhang and Green
(2025) this is an island of relatively higher $R_V$. In the outskirts
of the LMC, where we calibrate the TRGB, the $R_V$ and corresponding
$R_I$ reach rather lower values. Unfortunately, this cannot be
verified using the red clump stars because in these LMC regions of low
reddening, a measurement of the shift of the mean red clump magnitude
would be highly uncertain.
Thus, we assumed $R_I=1.34$ as our default value. Because we analyze fields
of low reddening in the LMC ($E(V-I) < 0.15$~mag) the uncertainty in
$R_I$ is, fortunately, a second order effect. Nevertheless, we
will show its impact on our results in Section 5.

The second correction to the RGB star brightness results from the
different distances of stars located in the LMC disk relative to the
observer. To calculate this correction we followed the standard flat
model of the LMC disk and used formulae presented in Section 3.4 of
Jacyszyn-Dobrzeniecka \etal (2016, see also van der Marel and Cioni
2001). For the assumed inclination, $i$, and the position angle of the
line of nodes, $PA$, we calculated the coefficients (a,b) of the disk
plane in the Cartesian system with the center located at the assumed
LMC center. Then, based on the Cartesian coordinates of an RGB star,
we calculated its real distance modulus (Eqs. 21, 22 in
Jacyszyn-Dobrzeniecka \etal 2016). The difference between the real
distance modulus and assumed distance modulus to the LMC center is the
second correction that has to be applied to the observed brightness of
an RGB star to place all stars at the same distance (\ie the distance
to the center of the Cartesian system).

In this study, we assumed the most accurate geometric distance to the
LMC from the OGLE eclipsing binaries (Pietrzy{\'n}ski \etal
2019). This distance was obtained assuming the LMC center at 
(RA, DEC) = (80\zdot\arcd05, $-69\zdot\arcd3$; van der Marel and
Kallivayalil 2014; HST+young population). We will later discuss
consequences of selecting a different center.

In each sector the {\it I}-band magnitude of the TRGB was extracted,
using a standard technique. First, a histogram of the RGB stars from a
sector was constructed in the range of $12.0 < I < 16.0$~mag with 0.02
mag wide bins. Then, the histogram was smoothed with the GLOESS
algorithm (Persson \etal 2004) using a $\sigma_s$ parameter equal to
0.125. Finally, an edge-detection Sobel-like filter with an extended
kernel ($[~-1,~-1,~-1,~0,~0,~0,~+1,~+1,~+1~]$; Madore, Mager and
Freedman 2009) was applied to the smoothed histogram, providing the
{\it I}-band magnitude of the TRGB. The typical accuracy of the TRGB
determination in each sector was about 0.020~mag. In the final step,
we combined the TRGB magnitudes from all 12 sectors and computed their
mean value and standard deviation, \ie their scatter.

\begin{figure}[htb]
\centerline{\includegraphics[width=8.1cm]{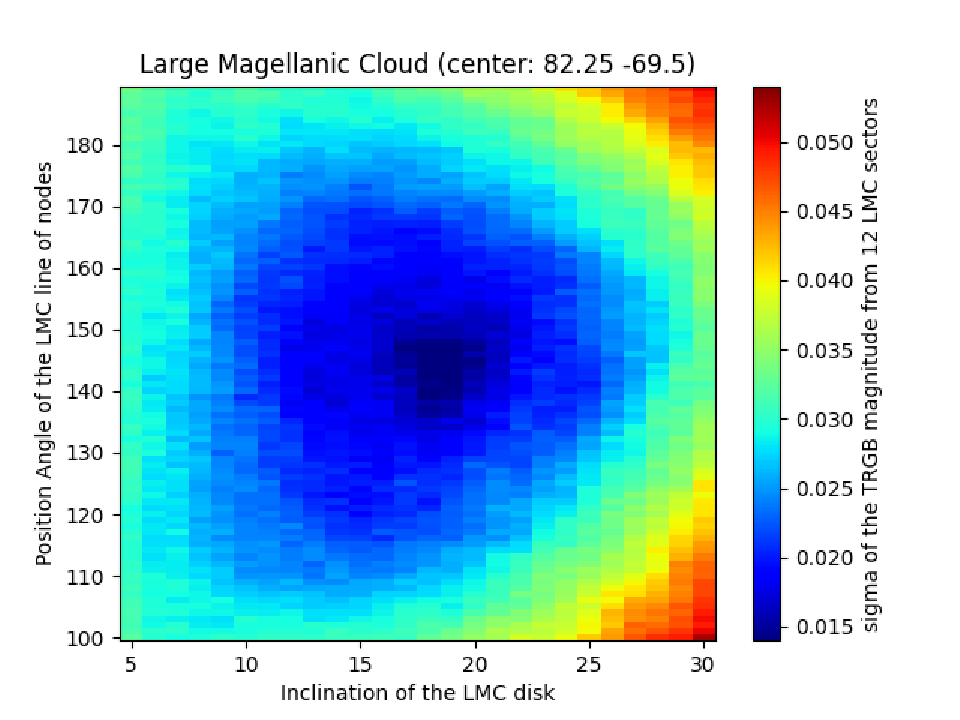}}
\vskip -5pt
\centerline{\includegraphics[width=8.1cm]{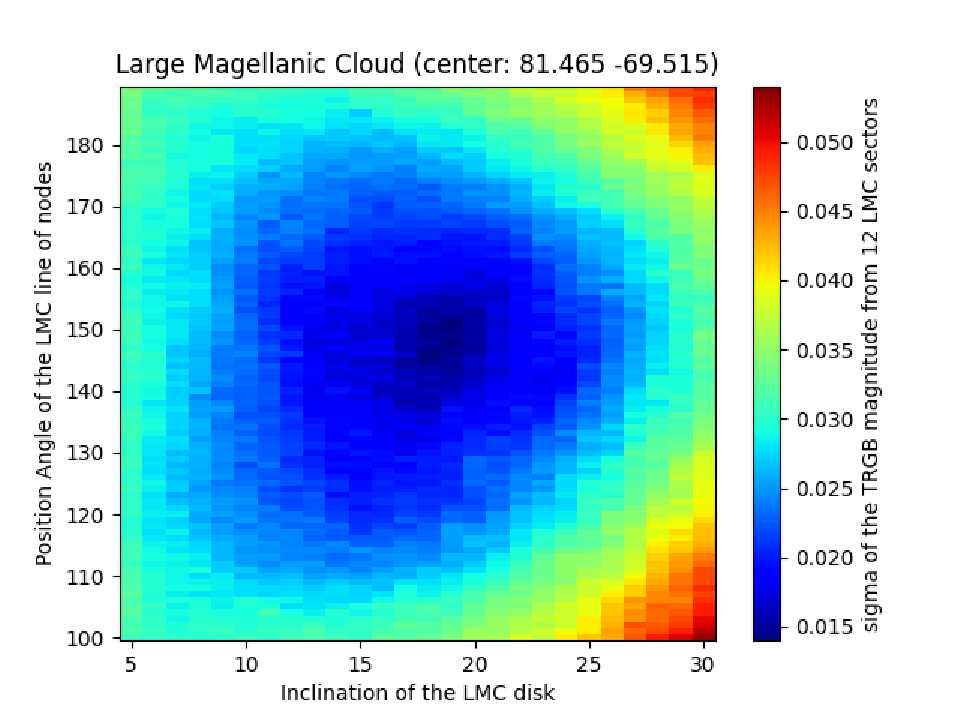}}     
\vskip -5pt
\centerline{\includegraphics[width=8.1cm]{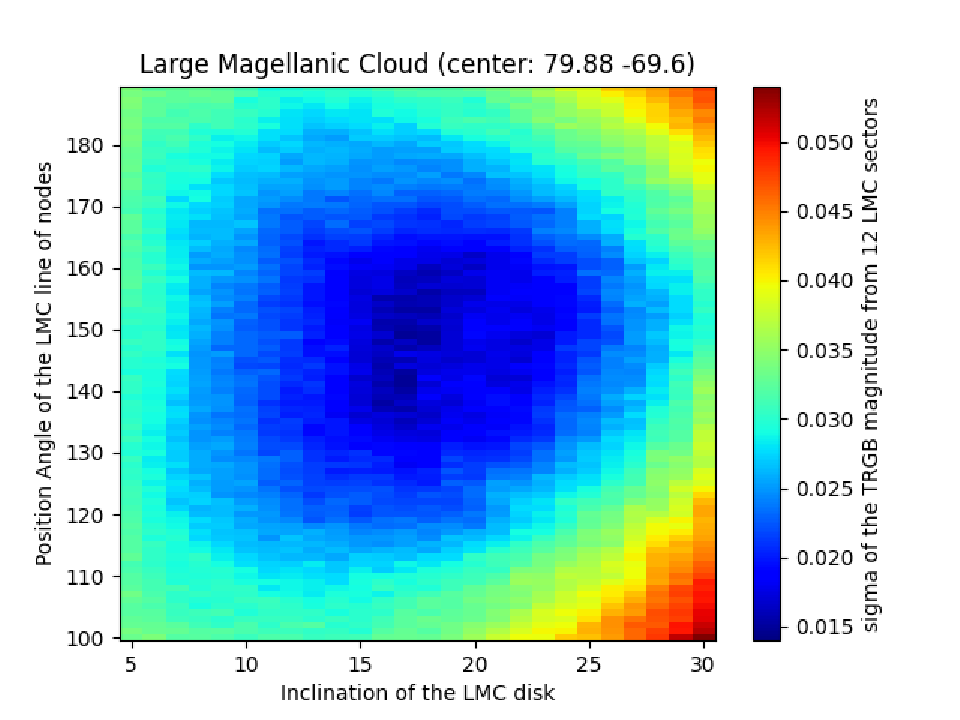}}
\vskip 5pt
\FigCap{The standard deviation of the {\it I}-band TRGB magnitude 
calculated from 12 sectors in the ($i$,$PA$) grid, assuming
$R_I=1.34$. Each of the three panels represents one of the assumed
centers of the LMC, as marked at the top pf the panel.}
\end{figure}


To verify how the TRGB magnitude we obtained depends on the disk
model parameters, we ran an iterative procedure on a grid of ($i$,
$PA$) LMC disk model parameters for three assumed LMC centers:
(82\zdot\arcd25, $-69\zdot\arcd5$; van der Marel and Cioni 2001),
(81\zdot\arcd465, $-69\zdot\arcd515$; Gaia Collaboration \etal 2021)
and (79\zdot\arcd88, $-69\zdot\arcd6$; van der Marel and Kallivayalil
2014). We varied the inclination in the range $5\arcd<i<39\arcd$, with
a step of one degree, and the position angle in the range 
$100\arcd< PA < 190\arcd$, also with a step of one degree. These
ranges cover, with a large margin, previous estimations of the LMC disk
parameters derived using different indicators. The standard deviation
(scatter) of the TRGB magnitudes from the 12 sectors served as an indicator
of the goodness of the disk model.

We also tested how sensitive the TRGB magnitude is to the $R_I$
parameter, by performing the above procedure for three values of $R_I$:
1.24, 1.34, and 1.44.

Fig.~3 shows the plot of the standard deviation of the TRGB magnitudes
from 12 sectors in the ($i$, $PA$) grid for each of the assumed
centers of the LMC and $R_I=1.34$. This {\it rms} is a measure of the
consistency of the TRGB magnitudes from all 12 sectors of the LMC disk
as a function of $i$ and $PA$. Our simulations indicate that for all
assumed LMC centers, the lowest standard deviation of the twelve TRGB
magnitudes occurs for disk inclinations $17\arcd<i<22\arcd$ and
position angles $140\arcd<PA<150\arcd$. This is very consistent with
previous estimations of the LMC disk parameters (\cf
Jacyszyn-Dobrzeniecka \etal 2016, Choi \etal 2018, Saroon and
Subramanian 2022), which is reassuring and shows that our geometric
corrections of a TRGB brightness are reliable.

In summary, our simulation consists of three parameters:
the LMC center (three variants), the segmenting of the LMC disk described
by $PA_0$ ($130\arcd$ and $150\arcd$), and the extinction parameter $R_I$
(1.24, 1.34, and 1.44). For each such simulation, we calculated the mean TRGB
magnitude from the 12 sectors. This was repeated for for disk inclinations
$5\arcd<i<39\arcd$ and  position angles $100\arcd<PA<190\arcd$.

Then, for each value of the three simulation parameters, we chose 25 values
of this mean TRGB magnitude for the ($i$, $PA$)
disk parameters for which the scatter, \ie the {\it rms} from averaging
the TRBG magnitude in these 12 sectors, was the lowest.
As a test of reliability, we also repeated these calculations using
the 250 best values instead of 25. In both cases, the results for the
mean TRGB magnitude of a given simulation were always consistent to
better than one mmag. The statistical error of the mean TRGB magnitude
of the simulation derived in this way was at less than one mmag level.

Table~1 summarizes the results of our initial calibration of the TRGB
with the LMC stars. For each simulation we provide the mean TRGB
magnitude from the 25 best LMC disk parameters (those having the
lowest scatter of TRGB magnitudes averaged over 12 sectors) and its
statistical uncertainty, as well as the typical scatter of the
individual TRGB magnitudes from the 12 sectors for these 25 best
determinations in a given simulation.

\renewcommand{\TableFont}{\scriptsize}

\MakeTable{cccccccc}{8.0cm}{The initial determination of the {\it I}-band
TRGB magnitude}
{\hline
\noalign{\vskip3pt}
&& \multicolumn{6}{c}{LMC  center} \\
$PA_0$ of& & \multicolumn{2}{c}{($79\zdot\arcd88$, $-69\zdot\arcd6$)}
& \multicolumn{2}{c}{($81\zdot\arcd465$, $-69\zdot\arcd515$)}
& \multicolumn{2}{c}{($82\zdot\arcd25$, $-69\zdot\arcd5$)}\\
A1, B1 & $R_I$ &\cline{1-6}\\
\noalign{\vskip-10pt}
& & Mean $I_0^{TRGB}$ & $rms_{12}$& Mean
$I_0^{TRGB}$ & $rms_{12}$ & Mean $I_0^{TRGB}$ & $rms_{12}$ \\
~[\arcd]& &[mag] &[mag] &[mag] &[mag] &[mag] &[mag] \\
\noalign{\vskip3pt}
\hline
\noalign{\vskip3pt}
130 & 		1.44		&	   $14.4518\pm0.0007$	& 0.016 &  $14.4459\pm0.0006$ & 0.017 &   $14.4432\pm0.0010$ &	0.017\\
150 & 		1.44		&	   $14.4547\pm0.0007$	& 0.024 &  $14.4402\pm0.0011$ & 0.018 &   $14.4464\pm0.0011$ &	0.019\\
& & & & & & & \\
130 & 		1.34            &          $14.4606\pm0.0008$   & 0.017 &  $14.4550\pm0.0009$ & 0.017 &   $14.4517\pm0.0007$ &	0.016\\
150 & 		1.34		&	   $14.4637\pm0.0008$  	& 0.024 &  $14.4492\pm0.0011$ & 0.018 &   $14.4550\pm0.0011$ &	0.019\\
& & & & & & & \\
130 & 		1.24            &          $14.4700\pm0.0009$  	& 0.017 &  $14.4632\pm0.0010$ & 0.017 &   $14.4617\pm0.0008$ &	0.017\\    
150 & 		1.24		&	   $14.4734\pm0.0012$  	& 0.024	&  $14.4581\pm0.0010$ & 0.018 &   $14.4647\pm0.0014$ &	0.021\\
\hline
}

%
%
%
%

\subsection{Final Determination of the {\it I}-band TRGB Brightness, $I_0^{TRGB}$}

Our tests performed in the previous subsection clearly indicate that
the upper RGB stars reside in the flat disk with parameters ($i$,
$PA$) in good agreement with other determinations based on a variety
of LMC structure tracers. More importantly, the {\it I}-band
magnitude of the TRGB very weakly depends on the disk parameters and
the location of the LMC center.

Keeping these results in mind, we decided to calculate the final
calibration of the TRGB based on the upper RGB stars from the outer
parts of the LMC, using the entire disk presented in Fig.~2 -- without
splitting it into sectors. Beside the already mentioned advantages of
working in these regions of the LMC, this approach has one more
significant benefit -- the number of stars for the TRGB analysis is
much larger, allowing a more precise determination of the TRGB. Within
our CMD limits for the upper RGB (see below) the number of RGB stars
from the entire LMC disk used in our analysis reaches 140\,000.

\begin{figure}[htb]
\centerline{\includegraphics[width=13cm, bb=10 45 550 630]{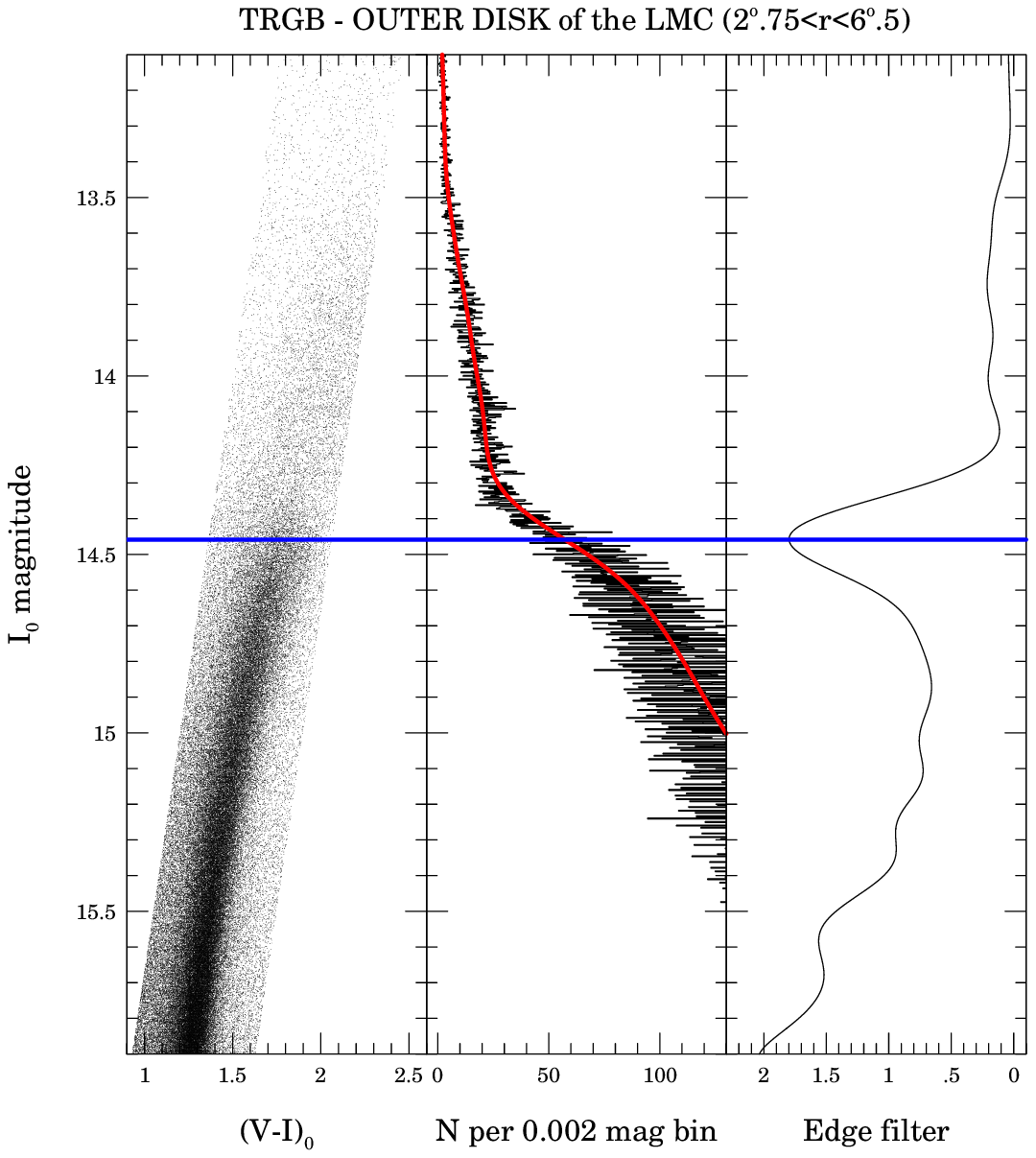}}
\vskip 12pt
\FigCap{Determination of the final TRGB brightness. {\it Left panel:}
Upper part of the RGB of the outer disk of the LMC in the $I_0$ \vs
$(V-I)_0$ CMD. {\it Middle panel:} Luminosity function in 0.002 mag
wide bins. The red line shows the GLOESS filter fit to the luminosity
function. {\it Right panel:} Sobel-like edge-detection function using
an extended kernel. Its maximum indicates $I_0^{TRGB}$ of the TRGB,
marked by the blue line across all panels.}
\end{figure}

As before, in the first step, for each individual star, we corrected
its observed magnitude and $(V-I)$ color for interstellar reddening
using the interstellar reddening map of Skowron \etal (2021) along a
given line-of-sight to obtain its intrinsic magnitude. We performed this
correction using three extinction parameters, $R_I$: 1.24, 1.34
(the most likely) and 1.44.

Then, we rectified the LMC disk by individually correcting the
brightness of stars located in different parts of the disk, as if they
were all at the same distance the from observer (\ie the distance to
the center of the Cartesian coordinate system -- see the previous
subsection). We assumed the disk parameters ($i$, $PA$) corresponding
to the best results from our initial tests -- the darkest blue points
in Fig.~3.

In the next step we followed the standard procedure for the TRGB
brightness determination. First, we constructed the CMD for the upper
part of the RGB limiting ourselves to stars brighter than $I_0<16$~mag
and located in the strip of 0.7~mag width in the $(V-I)_0$ color (from
0.9~mag to 1.6~mag at $I_0=16.0$~mag) and slanted with the slope of
$-3.3333$ (left panel in Fig.~4). Second, we created the luminosity
function (LF) of the upper RGB stars by binning them in 0.002~mag wide
bins (middle panel of Fig.~4). The LF was then smoothed using the
GLOESS filter (Persson \etal 2004) with the $\sigma_s$ parameter set
to 0.125. According to simulations carried out by Anderson \etal
(2024) this value lies in the middle of the optimal range of values
for this parameter for a typical upper RGB. The red thick line in the
middle panel of Fig.~4 shows the smoothed LF for the upper RGB of the
outer LMC disk.

There is a clear discontinuity in the LF corresponding to the TRGB. To
derive its precise magnitude we used the Sobel-like edge detection
algorithm. We applied the edge detection algorithm in a few flavors: using
the extended kernel: $[~-1,~-1,$ $~-1,~0,~0,~0,~+1,~+1,~+1~]$
(Madore \etal 2009) as the default; the standard Sobel kernel
$[~-1,~0,~+1~]$; the standard Sobel with signal-to-noise ratio (SNR)
weighting; and the standard Sobel with Poisson weighting, to double
check the results (\cf Appendix B1 of Anderson \etal 2024 and
references therein). Anderson \etal (2024) indicated that the two latter
weighting schemes introduce significant biases. The right panel of Fig.~4
presents the edge detection algorithm function for the extended
kernel. The thick blue horizontal line extending across all panels marks the
maximum of this function, \ie the $I_0^{TRGB}$ of the TRGB.

\renewcommand{\TableFont}{\footnotesize}

\MakeTable{cccc}{8.0cm}{Final determination of the TRGB {\it I}-band magnitude}
{\hline
\noalign{\vskip3pt}
& \multicolumn{3}{c}{LMC  center} \\
& ($79\zdot\arcd88$, $-69\zdot\arcd6$)
& ($81\zdot\arcd465$, $-69\zdot\arcd515$)
& ($82\zdot\arcd25$, $-69\zdot\arcd5$)\\
$R_I$ &\cline{1-3}\\
\noalign{\vskip-10pt}
& Mean $I_0^{TRGB}$ & Mean $I_0^{TRGB}$ & Mean $I_0^{TRGB}$ \\
&[mag] &[mag] &[mag] \\
\noalign{\vskip3pt}
\hline
\noalign{\vskip3pt}
1.44 &  $14.4491\pm0.0003$ &   $14.4431\pm0.0009$  &   $14.4405\pm0.0015$\\
{\bf 1.34} &  $14.4585\pm0.0005$ & ${\bf 14.4519}\,{\pmb \pm}\, {\bf 0.0004}$  &   $14.4491\pm0.0005$\\
1.24 &  $14.4684\pm0.0002$ &   $14.4622\pm0.0006$  &   $14.4596\pm0.0008$\\
\hline
}

We determined the $I_0^{TRGB}$ for the 25 best sets of the LMC disk
parameters ($i$, $PA$) from the simulations described in the previous
subsection (for $PA_0=130\arcd$ yielding lower {\it rms}), for each
pair of the parameters $R_I$ and the (RA, DEC) coordinates of the LMC
center.  We then averaged these values and calculated the $rms$
scatter of $I_0^{TRGB}$. The results of these calculations are listed
in Table~2 where the preferred value is written with bold font. To
verify the consistency of our results, we repeated these calculations
using the 250 best sets of ($i$, $PA$) disk parameters. The results
were consistent with those presented in Table~2 to within a fraction
of a mmag, with, as expected, a slightly larger $rms$. This again
demonstrates the very weak dependence of our result on the LMC disk
parameters.

To further test the robustness of our results we repeated our
calculations for two recent determinations of the red clump stars disk
parameters: Choi \etal (2018) -- ($i$, $PA$) = (27\zdot\arcd81,
146\zdot\arcd37) and Saroon and Subramanian (2022) -- (23\zdot\arcd26,
160\zdot\arcd43). Both measurements assume the LMC center at (RA, DEC)
= (82\zdot\arcd25, $-69\zdot\arcd5$; van der Marel and Cioni 2001). We
used $R_I=1.34$, allowing a direct comparison with the fourth column
and second row of Table~2, \ie $I_0^{TRGB}=14.4491\pm 0.0005$~mag. The
magnitudes of the TRGB, $I_0^{TRGB}$, for these two disk parameter
sets, are 14.4451~mag and 14.4452~mag, respectively. This is in
excellent agreement with our determination, which is highly
reassuring.

To summarize, the final values of the mean {\it I}-band TRGB
magnitude, $I_0^{TRGB}$, along with their $rms$ scatter from the 25
determinations with the best LMC disk parameters ($i$, $PA$), are
listed in Table~2 for three sets of the LMC center and three $R_I$
parameters. Our preferred value is highlighted in bold font. It is
striking that our ultimate determinations of $I_0^{TRGB}$ are in
excellent agreement with the results of our initial approach
(Table~1). This provides a strong confirmation of the reliability and
robustness of our results.

\subsection{$I_0^{TRGB}$ Uncertainty Analysis}

To assess potential sources of errors in our calibration we performed several
additional tests. First, we varied the LF binning scheme by adjusting the bin
width in the range of 0.001~mag to 0.020~mag, and by modifying the initial
bin position. A set of 25 such tests showed an {\it rms} scatter of
$I_0^{TRGB}$ at the 1~mmag level. 

Anderson \etal (2024) noted that the $\sigma_s$ parameter of the GLOESS
filter, used for smoothing of the RGB LF, should be carefully chosen and its
acceptable range depends on the shape of the RGB. Their simulations
indicate that for a typical RGB, such as that observed in the LMC, the optimal
range is $0.10<\sigma_s<0.15$. Our default value, $\sigma_s=0.125$ lies in
the middle of this range. We repeated calculations for the limiting values,
$\sigma_s=0.10$ and $\sigma_s=0.15$. Indeed, there is a weak dependence of
$I_0^{TRGB}$ on $\sigma_s$: for $\sigma_s=0.10$, $I_0^{TRGB}$ is 7 mmag
brighter and for $\sigma_s=0.15$, 6 mmag fainter. We included this possible
systematic uncertainty in the total error budget.

The second potential source of bias is the choice of the edge detection filter
and the method of calculating its response function: with of without weighting.
Anderson \etal (2024) noted that introducing weighting of the bins of the edge
function results in significant bias in the maximum of this
function, \ie the $I_0^{TRGB}$ magnitude. In particular, their
simulations show that the bias reaches 20 mmag when the SNR weights are used,
and even exceeds 60 mmag in the case of the Poisson error weights.
On the other hand, there is no bias when weighting is not used.
Therefore, for our final results, we decided to use pure (unweighted)
edge filter kernels,
in particular the extended one. However, both the standard and the
extended kernels yielded identical results. Nonetheless, we also
computed the edge function with SNR and Poisson error weighting schemes to
independently verify the predictions of Anderson \etal (2024). We found that indeed the
SNR weighting scheme yields an $I_0^{TRGB}$ value $\approx14$~mmag brighter, and
Poisson weighting scheme $\approx28$~mmag brighter. Given the TRGB contrast
parameter of our sample, $R=N_+/N_\_=4.41$, these biases are in a very good
agreement with their simulations (\cf Fig.~8 in Anderson \etal 2024).
In conclusion, our calibration of the TRGB does not use weighting for the
determination of the {\it I}-band magnitude of the TRGB. Thus, if used
for the precise Hubble constant determination, the TRGB magnitudes of
distant galaxies should also be derived without any weighting schemes to
avoid significant systematic errors.

The error budget of our TRGB magnitude, $I_0^{TRGB}$,
includes statistical errors -- the $rms$ of the mean TRGB magnitude
($\approx 1$ mmag), uncertainty due to the LF binning scheme
($\approx 1$~mmag), and the accuracy of the maximum of the edge function
determination ($\approx 5$ mmag). Systematic errors
include the accuracy of the OGLE {\it I}-band photometry zero point
($\approx 10$ mmag), uncertainty of the extinction parameter, $R_I$,
($\approx 10$~mmag) and $E(V-I)$ ($\approx 10$~mmag), uncertainty due to
the LMC center assumption ($\approx 5$~mmag), and the uncertainty resulting
from the choice of the $\sigma_s$ parameter during GLOESS LF smoothing
($\approx 7$~mmag).

As it is known, the magnitude of TRGB depends somewhat on metallicity of
red giants (Mager, Freedman and Madore 2008). Due to a small gradient
of metallicity with the distance from the LMC center observed in this
galaxy (Cioni 2009, Skowron \etal 2016) the observed RGB from the
outer disk of the LMC is a mixture of red giants of different
metallicities which may introduce some scatter of the TRGB
magnitude. Fortunately, the TRGB magnitude for the LMC $(V-I)_0$ color range in
this galaxy ($1.6<(V-I)_0<1.9$) is practically constant (Fig.~3 of
Mager \etal 2008). Nevertheless, we include an additional uncertainty
of 5 mmag due to this effect to the systematic error budget.     

Altogether, the final error budget of $I_0^{TRGB}$ is: $\pm 0.005~(\rm
stat.)$ and $\pm 0.020~(\rm syst.)$

\subsection{Absolute Magnitude, $M_{I,TRGB}$, of the TRGB}

It should be noted that the mean observed magnitudes of the TRGB,
$I_0^{TRGB}$, derived in the previous subsections correspond to the
distance of the assumed center of the LMC. To derive the absolute
magnitudes of the TRGB, we used the most precise geometric distance to
this galaxy (1\% accuracy) obtained from the OGLE eclipsing binaries
by Pietrzy{\'n}ski \etal (2019): $(m-M)=18.477\pm0.004~{\rm
(stat.)}\pm0.026~{\rm (syst.)}$~mag. This distance was derived
assuming the LMC center at (80\zdot\arcd05, $-69\zdot\arcd3$) and the
LMC disk parameters ($i=25\arcd$, $PA=132\arcd$). Thus, the distances
to the three LMC centers assumed in our simulations must be slightly
adjusted. In the bottom row of Table~3 we provide the distance moduli
to the center of the LMC adopted in this work, calculated based on
Pietrzy{\'n}ski \etal (2019) distance and their LMC disk parameters.
The final absolute magnitudes, $M_{I,TRGB}$, resulting from our
calibrations for each of our simulations are also listed in Table~3,
where the preferred value is in bold.

\MakeTable{cccc}{8.0cm}{Absolute {\it I}-band magnitude of TRGB}
{\hline
\noalign{\vskip3pt}
& \multicolumn{3}{c}{LMC  center} \\
& ($79\zdot\arcd88$, $-69\zdot\arcd6$) &
($81\zdot\arcd465$, $-69\zdot\arcd515$) & ($82\zdot\arcd25$, $-69\zdot\arcd5$)\\
$R_I$ &\cline{1-3}\\
\noalign{\vskip-10pt}
& $M_{I,TRGB}$& $M_{I,TRGB}$ & $M_{I,TRGB}$ \\
&[mag] &[mag] &[mag] \\
\noalign{\vskip3pt}
\hline
\noalign{\vskip3pt}
1.44		&	   $-4.033$   &   $-4.031$ &   $-4.031$ \\
{\bf 1.34}      &          $-4.024$   & ${\pmb -}{\bf 4.022 }$ &   $-4.022$ \\
1.24            &          $-4.014$   &   $-4.012$ &   $-4.011$ \\
\noalign{\vskip3pt}
\hline
\noalign{\vskip3pt}
\multicolumn{1}{l}{LMC $(m-M)$} & ~~18.482  & ~~18.474 & ~~18.471\\
\hline
}

\Section{Discussion}

\subsection{Absolute Calibration of the TRGB in the {\it I}-band based
on the LMC}

We presented here the {\it I}-band calibration of the TRGB based on
the red giant stars observed by the OGLE project during the OGLE-IV
phase in the outer parts of the LMC. We performed our determination of
the TRGB magnitude in two steps. In the initial approach we divided
the outer disk into segments to verify how well the LMC structure can
be approximated and how it affects the TRGB magnitude determination.

The final determination of the $I_0^{TRGB}$ TRGB magnitude is based on
the entire outer LMC disk. The additional advantage of this approach
is a much larger number of TRGB stars -- about 140\,000 in our sample,
than in individual segments, allowing the final determination to be
more precise.  We performed the $I_0^{TRGB}$ determinations for three
values of the extinction parameter, $R_I$, and three LMC center
positions (the same as in the initial determination). The results are
presented in Table~2.

For the ultimate calibration of the TRGB absolute magnitude in the
{\it I}-band we used the results from the final determinations of
$I_0^{TRGB}$. The distance to the each assumed center of the LMC was
appropriately adjusted. The ultimate calibration of the TRGB in the
{\it I}-band is presented in Table~3.

To summarize our calibrations, we conclude that the most likely mean
absolute magnitude of the TRGB stars in the LMC outer disk for our
default extinction parameter value of $R_I=1.34$ and the most likely
Gaia LMC center of TRGB stars (81\zdot\arcd465, $-69\zdot\arcd515$) is
$M_{I,TRGB}=-4.022$~mag. This result is highlighted in bold in
Table~3. It is highly unlikely that $R_I$ is larger than our default
value. If it is lower, our $M_{I,TRGB}$ may be a few mmag dimmer.

The main component of the error budget in our determination of
$M_{I,TRGB}$ comes from the uncertainty of the LMC distance
determination: statistical, $\pm0.004$ mag, and systematic,
$\pm0.026$~mag, (Pietrzy{\'n}ski \etal 2019). The remaining errors
resulting from our determination of the {\it I}-band TRGB magnitude of
the LMC, $I_0^{TRGB}$, are described at the end of Section~4.4. Thus, we
finally obtain the {\it I}-band absolute magnitude of the TRGB:

$$ M_{I,TRGB}=-4.022~\pm~0.006~{\rm(stat.)}~\pm~0.033~{\rm(syst.)}~{\rm mag} $$

It is clear that in order to further increase the precision of the
TRGB calibration, a more precise distance determination to the LMC is
needed.

\MakeTable{ccccc}{10.0cm}{OGLE absolute calibration of the TRGB from the LMC}
{\hline
\noalign{\vskip3pt}
$M_{I,TRGB}^{\rm No~bias}$ & $M_{I,TRGB}^{\rm SNR}$ & $M_{I,TRGB}^{\rm Poisson}$ &
$M_{I,TRGB}^{R=4, {\rm Poisson}}$ &  $M_{F814W,TRGB}^{\rm No~bias}$ \\ 
~[mag]& [mag]& [mag]& [mag] & [mag]\\
\noalign{\vskip3pt}
\hline
\noalign{\vskip3pt}
$-4.022$  & $-4.036$ & $-4.050$ & $-4.041$ & $-4.029$ \\
\noalign{\vskip3pt}
\hline
}

Table~4 shows our OGLE ultimate calibration of the {\it I}-band
brightness of the TRGB in different flavors -- ready for use in
various approaches to the Hubble constant determination. In addition
to our base, main value of $M_{I,TRGB}$ (interstellar extinction with
$R_I=1.34$; non-weighted edge filter, \ie no bias) we also list
calibrations for SNR-weighted edge filter, for Poisson-weighted edge
filter, for Poisson weighted edge filter reduced to the contrast ratio
of $R=4$ and, finally, our calibration for the HST F814W filter, which
closely resembles the {\it I}-band and has been widely used for
determining distances to galaxies with TRGB photometry from the HST
(using the {\it I} \vs F814W transformation from Freedman \etal 2020).

\subsection{Comparison and Calibration of the TRGB with the Small Magellanic Cloud}

The SMC is often used in TRGB calibrations 
as a benchmark for the consistency of obtained results. On the one
hand, it contains a large population of RGB stars, the interstellar
reddening -- foreground and intrinsic -- are generally lower than in
the LMC, the geometric distance to this galaxy has also been derived
with 2\% precision using OGLE eclipsing binaries (Graczyk \etal
2020). However, the main drawback of this galaxy is its highly
irregular and elongated shape along the line-of-sight. It is well known
that the eastern wing of the SMC is more than 0.5 mag closer in 
distance modulus than the western parts of the SMC
(Jacyszyn-Dobrzeniecka \etal 2016). Thus, the magnitude of TRGB
stars in the SMC is highly dependent on the line-of-sight,
making comparisons difficult. Additionally, due to the irregularities of
the SMC, its shape cannot be modeled as simply as a flat disk, as is
the case for the LMC.

Graczyk \etal (2020) were fully aware of the issue with the
unfavorable shape of the SMC and took great care in analyzing the
location of their OGLE eclipsing systems within the galaxy before
determining the average geometric distance to the SMC. Fig~4. in
Graczyk \etal (2020) shows that all but two (the closest and the
farthest) of the eclipsing systems used for the distance determination
(red dots) are located on the sky within a projected distance of 1~kpc
from the assumed SMC center (RA, DEC) = (12\zdot\arcd54,
$-73\zdot\arcd11$, Graczyk \etal 2020). This corresponds to
0\zdot\arcd9 on the sky. Therefore, we decided to calculate the TRGB
magnitude using RGB stars located within $r<0\zdot\arcd9$ from the SMC
center, expecting that their mean distance would coincide with the
mean distance of the eclipsing binaries. This is not an optimal region
within the SMC because it has relatively higher interstellar reddening
than the outer parts of this galaxy. However, this is not a major
issue, as the reddening in this region is still low -- comparable to
that in the outer parts of the LMC where we calibrated the TRGB. We
limited the SMC area used to the regions where $E(V-I)<0.15$~mag.

Within the radius of 0\zdot\arcd9 from the SMC center, we selected
about 20\ 000 upper RGB stars brighter than $I=16$~mag. Similar to
the LMC, each of these stars was treated individually and dereddened
using its line-of-sight reddening from the map of
Skowron \etal (2021). We assumed the extinction parameter $R_I=1.24$
as our default for the SMC (the $R_V$ parameter, and thus $R_I$,
seems to be somewhat lower in the SMC than in the LMC, Wang and Chen
2023). To double-check our calculations, we also repeated them
for $R_I=1.34$.

Next, we followed the standard procedure: we constructed a histogram of RGB stars
in 0.01~mag bins, smoothed it using the GLOESS filter with
$\sigma_s=0.125$ and applied the edge filter without a weighting scheme
to determine the TRGB magnitude.

To verify how our {\it I}-band TRGB magnitude depends on the size of
the assumed region in the SMC and whether this size is large enough to
smooth the irregularities near the center, we repeated our
calculations by selecting stars from a smaller region:
$r<0\zdot\arcd7$ from the center. This resulted in over 13\ 000 upper
RGB stars in our sample, which we binned in 0.02~mag bins to construct
the LF. The parameters of the GLOESS and edge filters remained the
same.

\MakeTable{ccccc}{8.0cm}{Tests of the OGLE TRGB calibration with the SMC }
{\hline
\noalign{\vskip3pt}
& \multicolumn{4}{c}{Small Magellanic Cloud (12\zdot\arcd54, $-73\zdot\arcd11$)} \\
& \multicolumn{2}{c}{$r<0\zdot\arcd9$}& \multicolumn{2}{c}{$r<0\zdot\arcd7$}\\
&\cline{1-4}\\
\noalign{\vskip-5pt}
$R_I$ & $I_0^{TRGB}$ & $(m-M)_{\rm SMC}$& $I_0^{TRGB}$ & $(m-M)_{\rm SMC}$\\
&[mag] &[mag] &[mag] &[mag]\\
\noalign{\vskip3pt}
\hline
\noalign{\vskip3pt}
{\bf 1.24}    & ${\bf 14.953}~{\pmb \pm}~{\bf 0.015}$ & ${\bf
18.975}~{\pmb \pm}~{\bf 0.015}~{\pmb \pm}~{\bf 0.033}$ & $14.949\pm0.018$ & $18.971\pm0.018\pm0.033$\\
& & & &\\
1.34    &  $14.948\pm0.015$ & $18.970\pm0.015\pm0.033$ & $14.941\pm0.018$ & $18.963\pm0.018\pm0.033$\\
\noalign{\vskip3pt}
\hline
}

Table~5 presents our results: the $I_0^{TRGB}$ magnitude and
$(m-M)_{\rm SMC}$ distance modulus obtained with our final TRGB
calibration. The statistical error in the TRGB magnitude determination
is 0.015~mag for the more numerous sample of $r<0\zdot\arcd9$ and
0.018~mag for $r<0\zdot\arcd7$. The systematic error in distance
modulus comes directly from the systematic error of our calibration,
primarily due to the uncertainty of the LMC distance.

Based on the OGLE eclipsing binaries Graczyk \etal (2020) obtained the
geometric distance modulus to the SMC equal to: $(m-M)_{\rm
SMC}=18.977\pm0.016~{\rm (stat.)}\pm0.026~{\rm (syst.)}$~mag, \ie with
$<2$\% accuracy. Our determination of the SMC distance modulus of 
$(m-M)_{\rm SMC}=18.975\pm0.015~{\rm (stat.)}\pm0.033~{\rm (syst.)}$~mag,
based on the TRGB stars in the same line-of-sight toward the SMC as the
eclipsing systems, is in remarkable agreement with the geometric
distance to this galaxy. Our TRGB distance is very little sensitive to
the assumed extinction parameter, $R_I$. On the other hand, our test
on the smaller area around the SMC center resulting in only a few mmag difference
in the TRGB brightness indicates that the $r<0\zdot\arcd9$ region of
the SMC, corresponding to the distribution of the eclipsing binaries
around the SMC center (Graczyk \etal 2020) is uniform enough for 
a precise TRGB distance determination.

The very good agreement between our TRGB distance modulus to the SMC
and the geometric distance modulus to this galaxy is reassuring,
confirming that our calibration of the {\it I}-band magnitude of the
TRGB is reliable and consistent with the most precise geometric
distance determinations to the LMC and SMC, with accuracies of 1\% and
2\%, respectively (Pietrzy{\'n}ski \etal 2019, Graczyk \etal 2020).

It should be noted that the contrast ratio of the TRGB for the
SMC fields is $R\approx3.75$. If the Poisson weighting scheme
of the edge detection function is applied, the $I_0^{TRGB}$ magnitude is
brighter by 33 mmags, and if the SNR weighting scheme is used, it is
brighter by 17 mmags compared to the $I_0^{TRGB}$ value in Table~5.
As in the LMC, both these
values are in good agreement with Anderson \etal (2024) simulations.

We may, of course, reverse the task and attempt to calibrate the {\it
I}-band absolute magnitude of the TRGB using the geometric distance to
the SMC (Graczyk \etal 2020). Taking the $I_0^{TRGB}$ for the region
$r<0\zdot\arcd9$ from the SMC center and the geometric distance to the
SMC (Graczyk \etal 2020) we obtain
$M_{I,TRGB}^{\rm No~bias}=-4.024~\pm0.22~{\rm (stat.)}~\pm0.26$~(syst.)~mag,
which is in perfect agreement with our outer LMC determination of
$-4.022\pm 0.006\pm0.033$~mag (Table~4).

\subsection{Comparison with the TRGB Calibration from NGC~4258}

Due to inconsistent absolute calibrations of the {\it I}-band
absolute magnitude of the TRGB based on the LMC photometry from different
studies, more attention has been given to the NGC~4258 galaxy. This
galaxy is the only extragalactic object, besides the Magellanic Clouds,
to which a precise geometric distance has been derived based on radio
observed water masers. Recent distance determinations to this galaxy
have reached 1.5\% accuracy (Reid, Pesce and Riess 2019), comparable to
the precision of the LMC distance. To derive the TRGB magnitude, this galaxy
must be observed from space, specifically by the HST or, more recently, by JWST.

Two recent attempts to precisely calibrate the {\it I}-band (or HST
F814W band) were undertaken by Jang \etal (2021) and Scolnic \etal
(2023). A comprehensive analysis of Jang \etal (2021) yields the F814W
filter magnitude of the TRGB: $M_{F814W,TRGB}^{\rm
SNR}=-4.050\pm0.028\pm0.048$~mag. To derive this value, Jang \etal
(2021) used the SNR weighting scheme of the edge filter function. This
brightness transforms to the {\it I}-band as: $M_{I,TRGB}^{\rm
SNR}=-4.043$~mag.
Our OGLE calibration of $M_{I,TRGB}$  with the same SNR weighting scheme is
: $M_{I,TRGB}^{OGLE, {\rm SNR}}=-4.036$~mag (14 mmag brighter than our {\it no bias}
value, see Table~4). This is in a very good agreement with Jang \etal
(2021) calibration of the TRGB with NGC~4258.

\MakeTable{cll}{10.0cm}{Comparison of the OGLE absolute {\it I}-band calibration of
the TRGB from the LMC with calibrations with NGC~4258}
{\hline
\noalign{\vskip3pt}
Calibrator& \multicolumn{1}{c}{$M_{I,TRGB}^{\rm SNR}$}&\multicolumn{1}{c}{$M_{I,TRGB}^{R=4,{\rm Poisson}}$} \\
& \multicolumn{1}{c}{[mag]}&\multicolumn{1}{c}{[mag]}\\
\noalign{\vskip3pt}
\hline
\noalign{\vskip3pt}
LMC (OGLE) & $-4.036~\pm~0.006~\pm~0.033$  & $-4.041~\pm~0.006~\pm~0.033$\\
NGC~4258   & $-4.043~\pm~0.028~\pm~0.048^a$ & $-4.030~\pm~0.035^b$\\
\noalign{\vskip3pt}
\hline
\noalign{\vskip3pt}
\multicolumn{3}{l}{$^a$Jang \etal (2021) -- CCHP}\\
\multicolumn{3}{l}{$^b$Scolnic \etal (2023) -- CATS}
}

Another calibration of the TRGB in NGC~4258 was performed by Scolnic \etal
(2023). Their analysis was based on the Li \etal (2023) data and took into
account the TRGB magnitude dependence on the contrast ratio, $R$, resulting
from the Poisson weighting of the edge detection function.
The {\it I}-band absolute magnitude of the TRGB, standardized to
$R=4$ yields $M_{I,TRGB}^{R=4, {\rm Poisson}}=-4.030\pm0.035$~mag.
This value can be compared with the OGLE Poisson weighted
$M_{I,TRGB}^{R=4, {\rm Poisson}}$ value in Table~4 equal to $-4.041$~mag
-- again, in a very good agreement.
The comparisons of our {\it I}-band calibration of the TRGB
in the LMC with those from the NGC~4258 galaxy are summarized in Table~6.

\subsection{Other Recent Calibrations of the {\it I}-band TRGB absolute magnitude} 

Another comprehensive analysis of the TRGB stars and the TRGB
calibration was performed by Hoyt (2023). This analysis, based again
on the OGLE-III photometry from the central parts of the LMC, yielded
$M_{I,TRGB}^{\rm Poisson}=-4.038\pm0.012\pm0.032$~mag. Comparing
with our result using an analogous edge detection function weighting scheme
(Table~4), namely $M_{I,TRGB}^{\rm Poisson}=-4.050$, we find that our
calibration is slightly brighter -- by about 12 mmags, assuming that
the contrast of the upper RGBs, $R$, is similar.

Finally, Anderson \etal (2024) presented another approach to calibrate
the TRGB by using OGLE small amplitude red giant (OSARG) variable
stars, also released by the OGLE-III project. Their final preferred
result -- the period-luminosity sequence B of the OSARG sample --
yields $M_{I,TRGB}^{\rm No~bias}=-4.018\pm0.014\pm0.033$~mag, in
excellent agreement with our result of $-4.022$~mag. However, their
all-red-giant sample gives a dimmer absolute magnitude of the TRGB,
namely $M_{I,TRGB}^{\rm No~bias}=-3.982\pm0.021\pm0.033$~mag.

\subsection{Impact of the New Calibration on the Hubble Constant Determination}

Although a full determination of the Hubble constant using our new
calibration is beyond the scope of this paper, we can assess how the
new ultimate calibration of the {\it I}-band absolute magnitude of the
TRGB would impact existing determinations of $H_0$ with the
TRGB. The final value of $H_0$ is quite a sensitive function of the
calibration magnitude. Freedman \etal (2020) noted that an increase of
$H_0$ by 4.4 km/s/Mpc would require a 0.13 mag fainter TRGB magnitude, \ie
10 mmag fainter magnitude increases $H_0$ by approximately 0.34
km/s/Mpc. Riess \etal (2022) estimated that a 10 mmag change in TRGB brightness
leads to a 0.33 km/s/Mpc shift in the Hubble constant. A similar estimate
of 0.38~km/s/Mpc was presented in Scolnic \etal (2023).

Our TRGB absolute magnitude calibration, which is more accurate than the
previous ones, differs by several mmag from those used in the recent TRGB-based
Hubble constant determinations. Thus, these existing values of $H_0$ can be
refined using our calibration.

In a series of papers (Freedman \etal 2019, 2020, Freedman 2021), the
CCHP project derived the final TRGB-based Hubble constant of
$H_0=69.8 \pm~0.6 \pm~1.6$ km/s/Mpc (Freedman 2021). An absolute magnitude
of $M_{F814W,TRGB}=-4.049\pm0.015\pm0.035$~mag,
corresponding to $M_{I,TRGB}=-4.042$~mag, was used for calibration in this
determination of $H_0$. However, it is not clearly stated which weighting scheme
of the edge detection function was used to derive the TRGB magnitude in the
individual components of this final calibration. As discussed earlier, the choice
of edge detection weighting can introduce a significant bias, which manifests as
a TRGB magnitude contrast relation (Anderson \etal 2024, Scolnic \etal 2023).
One of the main components of the Freedman's (2021) calibration is the
Jang \etal (2021) result based on NGC~4258 (Table~6). However, it uses
a signal-to-noise weighting scheme, in contrast to the Poisson weighting
applied to TRGB determinations in individual galaxies (Freedman \etal 2019).
It is also not clear which weighting scheme is used for the second main component
of the final calibration -- the OGLE-III LMC TRGB absolute magnitude (Freedman \etal 2020).

Assuming that, to be consistent with the TRGB magnitude measurements in
individual galaxies (which used the Poisson weighting), the Poisson weighted
variant of our new calibration -- $M_{I,TRGB}^{\rm Poisson}=-4.050$~mag, \ie
$M_{F814W,TRGB}^{\rm Poisson}=-4.057$~mag -- should also be used in the
comparison, then the resulting Hubble
constant would be lower by approximately 0.3~km/s/Mpc, giving $H_0=69.5$~km/s/Mpc.
However, keeping in mind the significant bias introduced by Poisson weighted
TRGB magnitudes as a function of the TRGB contrast, $R$, (Anderson \etal 2024),
the TRGB magnitudes of galaxies in the CCHP sample -- as well as our
calibration -- should first be corrected for this bias before a definitive
determination of $H_0$ from the CCHP dataset can be made.

Anand \etal (2022) presented another estimation of the Hubble constant
with the TRGB technique. They used the CCHP galaxy sample and carried
out an independent determination of the TRGB magnitude in the HST
F814W filter. However, contrary to the CCHP team, they employed a
completely different technique of determining the TRGB magnitude,
namely fitting the luminosity function of the asymptotic giant branch
(AGB) and RGB with a broken power law, where the break represents the
TRGB magnitude (Makarov \etal 2006). The TRGB magnitudes derived with
this method were then converted to distance moduli of galaxies using a
refined, HST $(F606W-F814W)$ color dependent calibration based on the
TRGB magnitude of NGC~4258 and its geometric distance modulus
(Anand \etal 2022, Eqs. 2 and 4), and then used for the determination
of $H_0=71.5\pm1.8$~km/s/Mpc.

However, the calibration used by Anand \etal (2022) predicts
unrealistic absolute magnitudes of the TRGB. For NGC~4258 with a color
of $(F606W-F814W)_0=1.32$~mag it yields $M_{F814W, TRGB}^{\rm
NGC~4258}=-3.993$~mag, \ie $M_{I, TRGB}^{\rm NGC~4258}=-3.986$~mag.
If applied to the LMC, where the TRGB has $(V-I)_0=1.8$~mag, corresponding to
$(F555W-F814W)_0\approx 1.9$~mag, and hence $(F606W-F814W)_0\approx
1.4$~mag (using transformations from Harris 2018, Jang and Lee 2017),
Anand's \etal calibration yields $M_{F814W, TRGB}^{\rm LMC}=-3.977$~mag, \ie
$M_{I,TRGB}^{\rm LMC}=-3.970$~mag. Given the geometric distance to the
LMC, the latter value implies an apparent magnitude of
$I_0^{TRGB}=14.507$~mag.  It is evident from Fig.~4 that this
brightness cannot be the real TRGB magnitude in the LMC.

The luminosity function technique applied by Anand \etal (2022) has
been shown to provide TRGB magnitudes consistent to within a few mmag
with those obtained with the standard GLOESS technique and an unbiased
edge filter (Kim \etal 2020, Table~2). Thus, it is justified to apply
our unbiased {\it I}-band magnitude of the TRGB, $M_{I,TRGB}^{\rm
No~bias}=-4.022$~mag (Table~4), to refine the Hubble constant derived
by Anand \etal (2022). The 52 mmag difference in absolute {\it I}-band
magnitudes of our LMC ultimate calibration and that resulting from the
Anand's \etal formula leads to a decrease of their Hubble constant
estimate by approximately 1.8 km/s/Mpc, resulting in
$H_0=69.7$~km/s/Mpc -- a value consistent with the Hubble constant
derived by the CCHP team.

Another attempt to determine the Hubble constant using TRGB distances
was undertaken by Scolnic \etal (2023) as a part of the CATS project.
They reanalyzed the HST observed galaxies, deriving Poisson weighted
TRGB magnitudes in the F814W filter, standardizing them to a contrast
of $R=4$, and calibrating distances using the TRGB magnitude
normalized to NGC~4258. Their resulting $H_0=73.2~\pm~2.0$~km/s/Mpc is
significantly different from the CCHP determination and is close to
the Hubble constant resulting from the Cepheid calibration.

Our Poisson weighted absolute calibration of the TRGB, normalized to $R=4$,
namely $M_{I,TRGB}^{R=4,{\rm Poisson}}=-4.041$, is 11 mmag brighter than
that used by Scolnic \etal (2023) (Table~6). Thus, applying our
calibration to the CATS sample, the Hubble constant would be smaller by 0.4
km/s/Mpc, \ie $H_0=72.8$~km/s/Mpc, still in very good agreement
with the value from the Cepheid-based $H_0$ determination (Riess \etal
2022). It should be noted, however, that the accuracy of distances to
galaxies from the CATS sample has recently been questioned by Hoyt \etal
(2025), so it seems that the reliability of this $H_0$ determination
still remains an open question.

In summary, refining the Hubble constant determinations with the TRGB
magnitude based distances and our new ultimate TRGB calibration
indicates that the local Hubble tension still persists. The CCHP
result remains practically the same, Anand \etal (2022) determination
(based on a different TRGB determination technique) becomes consistent
with the CCHP value, and the status of the Scolnic \etal (2023)
determination remains unclear. We conclude that this tension is not
caused by an uncertain calibration of the absolute {\it I}-band
magnitude of the TRGB, as our ultimate calibration is in perfect
agreement with all three most precise geometric distance determination
to the LMC, SMC and NGC~4258. The discussion of other possible sources
of the local Hubble tension is beyond the scope of this paper.

\Section{Summary}

In this paper, we presented the ultimate {\it I}-band absolute
magnitude of the TRGB:

$$ M_{I,TRGB}=-4.022~\pm~0.006~{\rm(stat.)}~\pm~0.033~{\rm(syst.)}~{\rm mag} $$
which may serve as the basis for calibrating the extragalactic distance
measurements using the TRGB technique and for determining the
Hubble constant. Various versions of this calibration for different TRGB
applications are listed in Table~4.

Our TRGB calibration is based on a precise analysis of the outer regions
of the LMC -- areas well suited for such a study due to low
interstellar reddening, lower stellar crowding, and the simple modeling of
the LMC structure in these regions.  The accuracy of our TRGB absolute
magnitude is limited primarily by the 1\% uncertainty in the geometric
distance to the LMC, and can therefore be refined in the future.

We verified the precision of our TRGB calibration by analyzing two
galaxies with geometric distances known to percent-level accuracy,
namely the SMC and NGC~4258, and found excellent agreement. We
also attempted to refine existing determinations of the Hubble
constant using TRGB distances, and concluded that the local Hubble
tension -- the discrepancy between $H_0$ values derived from TRGB and 
Cepheid distances -- still persists. It is, however, not caused by the absolute
calibration of the TRGB, which is now very robust.

Significant progress in determining the Hubble constant and resolving
both the main and local Hubble tensions is expected from much deeper
observations of distant galaxies with the JWST telescope
(Freedman \etal 2025, Li \etal 2024). Unfortunately, the JWST filter
system does not include a band closely resembling the {\it
I}-band. The closest available filter is F090W, along with a full set
of near-infrared filters. However, the TRGB is not as effective a
standard candle in these bands as it is in the {\it I}-band.  Thus,
additional new techniques of correcting the near-infrared TRGB, as
well as new absolute calibrations, must be developed to fully exploit
the huge potential of JWST (\eg Hoyt \etal 2025, Anand \etal
2024). One should be aware, however, that this is always a potential
source of additional uncertainty.

\Section{Data Availability}

The OGLE-IV deep photometry of the outer regions of the Large Magellanic
Cloud and the central region of the Small Magellanic Cloud, along with the
interstellar reddening data (Skowron \etal 2021) used in this paper, are
available through the OGLE website:

\begin{itemize}
\vspace*{-7pt}
\parskip=0pt \itemsep=1mm \setlength{\itemsep}{0.4mm}
\item {\it https://ogle.astrouw.edu.pl/cont/4\_main/map/} -- the OGLE Web page
\item {\it https://www.astrouw.edu.pl/ogle/ogle4/TRGB} -- the OGLE Internet archive
\end{itemize}

\Acknow{We would like to thank Drs.\ Barry Madore and Taylor Hoyt for
sharing with us the GLOESS Python code.}

\end{document}